\documentclass[aps,prl,twocolumn,showpacs,superscriptaddress,groupedaddress,nofootinbib]{revtex4-1}  
\usepackage{graphicx}  
\usepackage{dcolumn}   
\usepackage{bm}        
\usepackage{amssymb}   
\def\be{\begin{equation}}
\def\t{\tau}
\def\s{\sigma}
\def\ee{\end{equation}}
\def\be{\begin{equation}}
\def\ee{\end{equation}}
\def\sl2{$SL(2,\mathbb{R})$}
\begin{document}
\title{Horizon Strings and Interior States of a Black Hole}
\author {K. P. Yogendran}
\affiliation{IISER Mohali (on deputation to IISER Tirupati) }
\date{\today}

\begin{abstract}
  We provide an explicit construction of classical strings that have
  endpoints on the horizons of the 2D Lorentzian black hole. We argue
  that this is a dual description of geodesics that are localized
  around the horizon which are the Lorentzian counterparts of the
  winding strings of the Euclidean black hole (the cigar
  geometry). Identifying these with the states of the black hole, we
  can expect that issues of black hole information loss can be posed
  sharply in terms of a fully quantizable string theory.
\end{abstract}

\maketitle


Any resolution of the black hole information paradox requires some
account of the interior of the black hole. Stating the question thus
carries the implication that computations of the entropy of a black
hole, which rely on a microcanonical ensemble description valid at
$G_N=0$ (such as the degeneracy counting based on D-brane models) will
not suffice to address unitarity of quantum processes at the horizon
since the horizon vanishes at $G_N=0$. An approach using perturbative
string theory could conceivably lead to novel insights, but this
requires the knowledge of the complete string spectrum on a black hole
background which has not yet made an appearance.

In \cite{Susskind}, Susskind has suggested that the entropy of the
black hole be modelled in terms of strings that end on the
horizon. Such strings will be accessible to the external observer and
to the interior observer alike and will presumably be implicated in a
unitarity description of black hole evaporation etc. It will therefore
be of much interest if such strings can be constructed explicitly and
perhaps even quantized.

In this context, there are two exactly solvable sigma models whose
target space interpretation is that of a black hole: the BTZ black
hole in 3D \cite{BTZ} and the 2D black hole \cite{Witten} (one can
construct other exact sigma models by building on these CFTs). The
presence of a trapped region bounded by a horizon enables one to
formulate all the usual black hole puzzles. It is therefore
interesting to explore the spectrum of these models to isolate string
configurations that might be relevant in studies of the information
paradox and black hole complementarity.

The AdS/CFT correspondence has been a significant guiding force in
recent studies of these issues and suggests that the interior of the
black hole must be assigned a Hilbert space of states. Several studies
have attempted to identify correlators of the dual field theory that
would involve these interior modes and hence provide insights into the
quantum black hole.  The recent ER$=$EPR idea \cite{EREPR} suggests
that one should search for Einstein-Rosen bridge like configurations
as being the source of the Bekenstein-Hawking entropy which is now to
be thought of as entanglement entropy between the two asymptotically
flat regions of the extended black hole geometry.

For both the 2D black hole and the BTZ geometry, some version of AdS
holography presumably exists but has not been completely
identified. Thus, a construction of the full quantum spectrum of
string theory on these backgrounds can be expected to be a fertile
playground for the exploration of issues relating to the information
paradox and black hole complementarity.

In this letter, we shall focus on the 2D black hole and identify
classsical configurations which possess the various desiderata as
discussed above, and are also quantized relatively easily. 

{\bf The 2D black hole as a gauged sigma model}

We shall present a quick summary of the sigma model describing strings
propagating on the Lorentzian 2D black hole as a gauged conformal
field theory \cite{Witten}. This black hole has a dual description as
a condensate of non-singlet modes of a matrix model \cite{KKK}
and was in fact, first constructed as a solution to the low energy
effective action\cite{Mandal}.

To begin with, we start with the WZNW model based on the group \sl2
which is a noncompact conformal field theory. The quantum spectrum of
(the universal cover) of this model was fully realized in the work of
Maldacena and Ooguri \cite{MO1} who showed that a modular invariant
partition function maybe obtained upon including spectrally flowed
representations of the current algebra.

The symmetry that is gauged corresponds to a hyperbolic subgroup of
\sl2 generated by the current $J^3-\bar J^3$, which acts on
\be
SL(2,\mathbb{R})\ni g = \left(\begin{array}{cc} a &
  u\\-v&b \end{array} \right)  \to
g'=e^{\epsilon\sigma_3}\,g\, e^{\epsilon\sigma_3}\label{axial}
\ee
where $\epsilon(\s,t)$ is the gauge parameter. 

\begin{figure}
\includegraphics[scale=0.4]{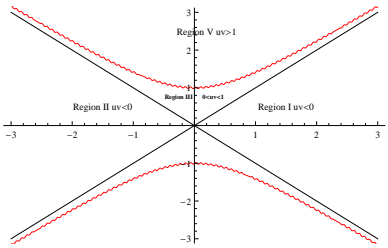}
\caption{\label{charts} The extended Lorentzian black hole
 geometry: The diagonal lines are the horizons while the red
 hyperbolae are the spacelike singularities}
\end{figure}

The target space interpretation of the coset theory maybe obtained by
gauge fixing and integrating out the gauge fields. The resultant sigma
model has a metric and a dilaton \[ds^2= -k \frac{du dv}{1-uv} \qquad
\Phi=-\frac{1}{2}\log (1-uv)\] where the off-diagonal entries of the
\sl2 matrix descend as embedding coordinates. This metric is that of a
black hole with horizon represented by the diagonal lines $uv=0$ (see
Fig.\ref{charts}), while there is a curvature singularity at
$uv=1$.

Early studies of this string theory \cite{DVV,DistNel} showed that the
spectrum consisted of a single massless scalar field, together with
the possibility of massive states at discrete values of the momenta.
In view of the necessity of spectrally flowed representations for the
parent \sl2 theory, it is of interest to revisit the spectrum of the
black hole. In this context, we note that the Euclidean black hole
which has the shape of a cigar (and is itself another gauged CFT) has
winding strings in its spectrum which can be shown to arise from these
spectrally flowed representations.

We can approach the spectrum by studying classical particle-like
solutions to the sigma model (for earlier work along these lines, see
\cite{Bars}). In this two dimensional case, these are likely to be
the whole story whereas in higher dimensions, we can have strings,
D-branes and other extended objects in the spectrum. Particle like
modes of the string, in the classical limit, will propagate on
geodesics of the black hole geometry which are easily constructed. As
an example, a null geodesic of the black hole is $$ u= -1 \quad
v=(e^{2E\tau}-1), $$ with $E$ the energy, which we will regard as a
possible solution of the black hole sigma model. We can then uplift
this to the WZNW model as an \sl2 matrix \be g=\left(\begin{array}{cc}
  e^{E(\tau-\sigma)}&-1\\1-e^{2E\tau}&\,e^{E(\tau+\sigma)}
\end{array}\right).\label{null-lift}\ee
Because of the gauge symmetry, the diagonal entries are ambiguous and we have
used this to introduce $\sigma$ dependence in the diagonal entries. It
is now easy to see that this matrix can be factorized into a product
of left and right movers \[
g=e^{-\frac{E\sigma^+}{2}\sigma_3}\,g_+(\sigma^+)g_-(\sigma^-)
e^{\frac{E\sigma^-}{2}\sigma_3}\] and hence is a solution of the \sl2
WZNW model. Further, it can be shown that the product
$g_+(\sigma^+)g_-(\sigma^-)$ represents a {\em spacelike} geodesic of
\sl2 and hence upon quantization, will be a state in the principal
continuous series of \sl2. The additional factors of $e^{\pm
  \frac{E\sigma^\pm}{2}\sigma_3}$ is the operation of spectral flow.
Thus the null geodesic of the black hole is obtained from the
{\em spacelike} geodesic of \sl2 after spectral flow along the
$\sigma_3$-direction in \sl2. The $J^3$ current of \sl2 and the
worldsheet stress tensor undergo shifts
\be J^3_\pm =\tilde J^3_\pm +
k \frac{E}{2} \quad T_\pm=\tilde T_\pm+ E \tilde J^3_\pm +
k\frac{E^2}{4}. \ee
under the spectral flow operation where $\tilde T_\pm, \tilde J^3$
refer to the currents of the \sl2 CFT.

We may similarly uplift the timelike geodesics of the black hole
geometry as \sl2 matrices. The timelike geodesics come in three
classes depending on the sign of $E^2-m^2$. In the case $E^2>m^2$, the
geodesics of the black hole scatter out to asymptotic infinity and can
be shown to be again obtained from spacelike geodesics of \sl2 after
the spectral flow operation. Thus, the null geodesics and timelike
geodesics with $E^2>m^2$ are both obtained from the principal
continuous series of \sl2, and as such can be mapped into each other
by an \sl2 transformation (before spectral flow). This is probably the
origin of the duality observed in \cite{DistNel} which interchanges
the massive and the massless states of the black hole theory.

\begin{figure}
  \includegraphics[scale=0.7]{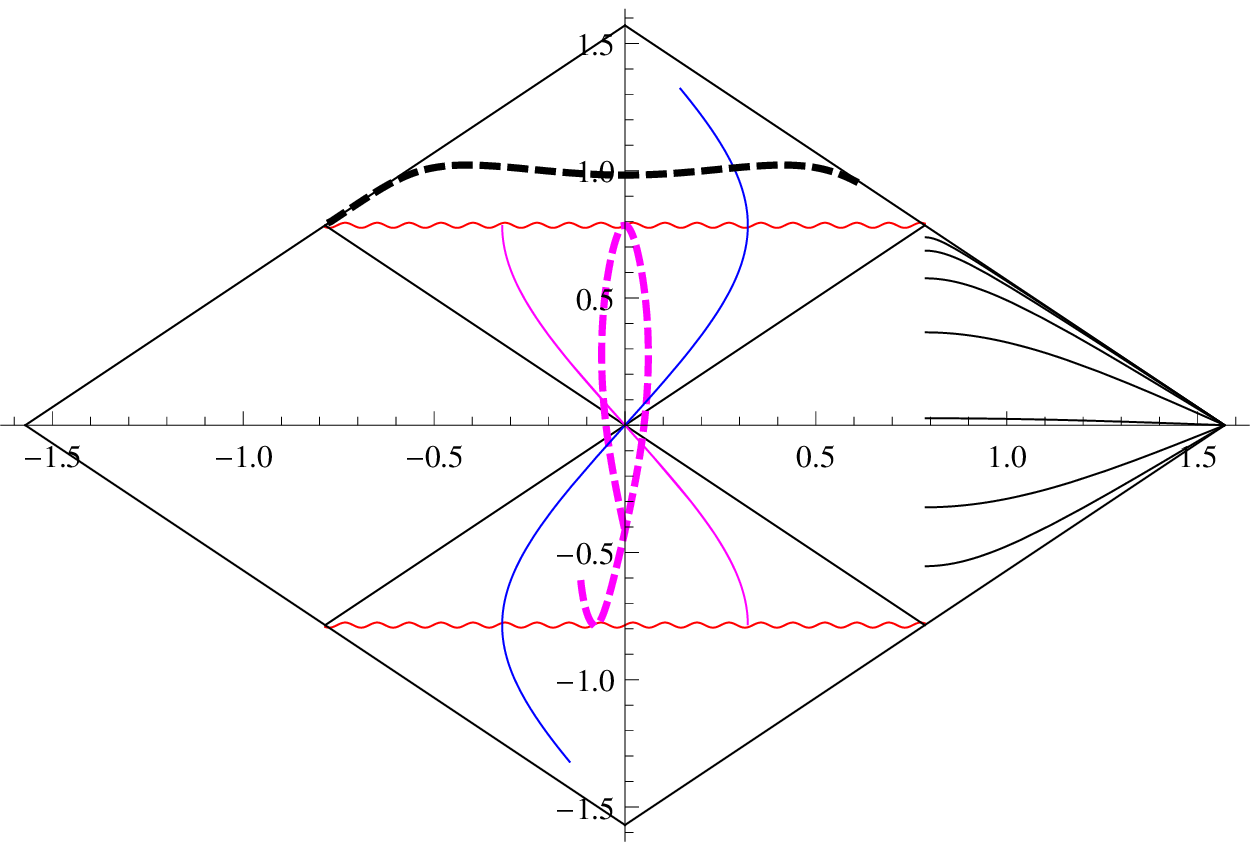}
\caption{\label{Fig.states} The timelike geodesics and the new string modes}
drawn on the Penrose diagram
\end{figure}

We shall focus our attention on the geodesics with $E^2<m^2$.  These are all
localized around the black hole horizon and never reach asymptotic infinity
(the dashed magenta curve in the middle in Fig.\ref{Fig.states}). 
For e.g.,
\[u=-\frac{e^{-E\tau}}{\sin\phi}\sin(\beta\tau+\phi)\quad
v=\frac{e^{E\tau}}{\sin\phi}\sin(\beta\tau-\phi)\] which is seen to
satisfy $uv>-\cot^2\phi$ i.e., it never reaches the asymptotically
flat region at $uv\to -\infty$.  Here the parameters $\beta^2=m^2-E^2$
and $\tan^2\phi=\frac{\beta^2}{E^2}$. When uplifted to \sl2, this
geodesic is represented by \be
g=\frac{1}{\sin\phi}\left(\begin{array}{cc} e^{-E\sigma}\sin\beta \tau
  &
  -e^{-E\tau}\sin(\beta\tau+\phi)\\ e^{E\tau}\sin(\beta\tau-\phi)&e^{E\sigma}\sin\beta\tau
\end{array}\right).\label{local-lift}
\ee
Here, we will regard the parameters $\beta$ and $\phi$ as being
independent of $E$, and we may expect that the level matching and
Virasoro conditions of the string theory determine the relations
between $\beta,E,\phi$. It can be shown that this matrix is obtained
from the {\em time-like} geodesic of \sl2 by applying the spectral flow
operation, i.e.,
$g=e^{-E\sigma^+\sigma_3}\,U\,e^{i\sigma_2\tau}\,V\,e^{E\sigma^-\sigma_3}$
($U,V$ are constant \sl2 matrices). Hence, upon quantization, these
geodesics will give states in the Discrete Series of \sl2. It can now
be shown that these geodesics, upon ``Wick rotation'', map
to the winding strings of the Euclidean black hole.
Hence, we propose that these should be interpreted as the single
particle internal energy levels of the black hole. Since the quantum
wavefunctions corresponding to these trajectories will have vanishing
norm at asymptotic infinity, this supports their interpretation as
``bound states'' of the geometry. This interpretation will be
justified by a dual description below. We also note that the geodesics
extend into both regions I and II through the interior (see
Fig.\ref{charts}) of the black hole geometry which means that the
boundary quantum states corresponding to these are entangled (across
the horizon). Thus, a count of such quantum strings should account for
the Bekenstein-Hawking entropy, now viewed as the entanglement entropy
of the left and right black holes \cite{EREPR}.

{\bf T-duality and New states}

In the usual description of T-duality, which is a symmetry of string
perturbation theory, one either performs a spacetime $R\to
\frac{1}{R}$ transformation or a worldsheet $\tau\to \sigma$
transformation. Either way, we end up with the same string moving on a
different target space (with radius $1/R$). If, however, we perform
both transformations, then a string state $|n,w\rangle$ with momentum
and winding quantum numbers $n,w$ is transformed into a different
physical state with quantum numbers $|w,n\rangle$. A similar
transformation can be performed for this black hole sigma model as
well, and will give us new, T-dual extended string configurations
corresponding to each geodesic of the previous section.

In this case, the analogue of T-duality is the axial-vector
interchange of the symmetry that is being gauged.  A special feature
of the black hole sigma model is that it is self dual under this
interchange, that is if the symmetry we gauge is vectorial $g\to
g'=\exp(\epsilon\sigma_3)\,g\exp(-\epsilon\sigma_3)$, then the
resultant coset sigma model turns out to have the same target space as
in the former case Eq.\ref{axial}. In this case, the diagonal entries
of the \sl2 matrix are gauge invariant and form the coordinates for
the extended black hole spacetime. Since $ab=1-uv$, the asymptotically
flat region I ($uv<0, ab>1$) in front of the horizon is dual to the
region V ($uv>1,ab<0$) on the other side of the singularity.

We can therefore consider one of the matrices that represent a geodesic of
the black hole geometry, and now regard the entries $a(\tau,\sigma)$
and $b(\tau,\sigma)$ as the embedding coordinates of a string
worldsheet in the dual region. This is equivalent to right
multiplication by $i\s_2$, and hence will also give a classical solution
of the \sl2 theory. This will however result in flipping the right
moving $\bar J^3$ quantum number which will not be compatible with the
level matching condition of the axially gauged theory (unless the
quantum number is zero).

If, in addition to right multiplication by $i\s_2$, we perform the
worldsheet $\tau\to\sigma$ transformation, the resultant worldsheet
will be a classical solution of the original string theory. Now
the $\bar J^3$ changes sign twice and hence, if the original solution
was physical, the dual will also be physical.

For instance, in the matrix (\ref{null-lift}) representing a null
geodesic in region I, $a=\exp^{E(\tau-\s)}\quad b=\exp^{E(\t+\s)}$
from which, after interchanging $\tau\to\s$ and reinterpreting, we get
$uv=e^{2E\s} \quad t=-E\tau$ which we expect to be a solution in
region V (and III). If we start with a timelike geodesic in region V
(the horizontal black dashed curve in Fig.\ref{Fig.states}), upon
dualising we get folded strings
$uv=-\frac{\cosh^2\beta\s}{\sinh^2\phi}$ that reach out to the
boundary at $uv\to -\infty$, but do not reach the horizon at
$uv=0$. The black curves ending at the right boundary in
Fig.\ref{Fig.states} represent this folded string at a few instants of
worldsheet time.  We should, using a holographic interpretation,
regard these as being dual to operators of the boundary theory living
at $uv\to -\infty$. Thus, we obtain ``boundary operators'' from the
particle trajectories of region V upon dualization.

{\bf Horizon Strings}

However, the geodesics with $E^2<m^2$ (belonging to the Discrete
series) are the most interesting of them all. Firstly, considering the
$a,b$ entries from Eq.\ref{local-lift} as representing a solution of
the {\em vectorially} gauged theory, we can show that these represent
the Lorentzian version of the ``winding modes'' of the trumpet
geometry (which is region V upon Wick rotation). This is as expected
{\em if} the original modes were the winding modes of the cigar since
the trumpet geometry is T-dual to the cigar. In \cite{KKK}, these are
the modes that condense in the Sine-Liouville description to form the
black hole.

By multiplying the matrix in Eq.\ref{local-lift} by $i\s_2$ and
interchanging $\t,\s$, we get a new solution of the original theory
\[u=\frac{\sin\beta\s}{\sin\phi} e^{-E\t} \quad v=\frac{\sin\beta\s}{\sin\phi}e^{E\t}\]
which is always inside the horizon but extends across the singularity
($0\leq uv \leq \rm{cosec}^2\phi$). In Fig.\ref{Fig.states}, the solid vertical
curves in the middle in magenta and blue are two such representative
configurations, drawn at fixed $\t$.  These worldsheets are ``stuck''
to the horizon at $\s=0$ and folded over at $\s=0,2\pi$ (since we
might expect that $\beta\in \mathbb{Z}$), and thus meson-like (to use
the holographic terminology). This is interesting because one might
wish to regard the black hole as being a condensate of these
``mesons'' consistent with a holographic interpretation of the finite
temperature state of the dual theory.

Thus, we might choose to regard the black hole as a condensate of
``mesons'' as above. Or else, regard the ``bound'' geodesics as the
degrees of freedom of the black hole visible to the external
observer. These are complementary to each other in the sense of
vector-axial duality and also possibly complementary to each other in
the sense of black hole complementarity because the interior modes are
visible only to an observer who has crossed the horizon. Note that
both the geodesic and the folded string description are simultaneously
applicable only in the interior of the black hole geometry. Also, in
region V (which is the Lorentzian version of the trumpet geometry), only
the folded string interpretation exists, there are no geodesics with
$E^2<m^2$.


{\bf Level Matching and Physical State Condition}

For a classical solution to represent a physical state of the string
theory, we need to impose the Virasoro and level matching
conditions. The latter, in the case of axial gauging, is $J^3=\bar
J^3$. All the geodesic solutions that we have constructed satisfy this
condition.

Upon applying either the $\t\to \s$ transformation or the $i\s_2$
multiplication, $\bar J^3$ changes sign and hence the equation cannot
be met unless the $J^3=\bar J^3=0$.  But if both operations are
performed and if the original configuration satisfies the level
matching condition, then the resultant configuration will also satisfy
$J^3=\bar J^3$.

The Virasoro condition $ T_{bh}\equiv T_{sl_2}-\frac{1}{k} (J^3)^2=0
\quad $ however imposes conditions on the various quantum numbers. For
instance, in the case of null geodesics, we obtain the dispersion
relation $\beta^2=E^2$. Classically, the massive geodesics all do not
satisfy the Virasoro conditions.  Upon quantization and following the
analysis in \cite{DVV,DistNel}, we can obtain massive states for
discrete values of the quantum numbers (the difference with that work
being the inclusion of spectral flow).






{\bf Discussion}

To summarise, we have argued that the string spectrum requires
spectrally flowed current algebra representations based on the
Discrete series as well as the Continuous series of \sl2.  This
construction of the full Hilbert space of the string theory has some
attractive features which we would like to highlight. Although the
above discussion was entirely with reference to the 2D black hole,
much of it carries over, mutatis mutandis, to the BTZ black hole (and
is already present in the literature \cite{EKVH}). Thus, one can
compare calculations (at least those that pertain to the information
paradox) in these two systems.

Upon quantization, the states of the Continuous series can be shown to
be doubled in a manner that reflects the presence of the two
asymptotically flat regions.  This doubling, which is required by the
representation theory of \sl2, leads to entanglement of regions I and
II (Fig.\ref{charts}) of the black hole.  The operators of the
Discrete and Continuous series are then represented as 2-d matrices
(besides the vertex operator part) acting on both copies of the
Hilbert space \cite{Mukunda}.
The curious identity remarked in \cite{Giveon} following earlier work,
suggests that it is necessary to include the Discrete series in the
spectrum to achieve modular invariance (we have not established
necessity). It is of course important to check that ghosts are absent
in the spectrum of the string theory, and establish modular invariance
in the hyperbolic basis that is necessitated by the black hole
interpretation.  The localised geodesics which extend across the
horizon appear to be the right configurations to create the entangled
state represented by the black hole along the lines of the ER$=$EPR
proposal of Maldacena and Susskind \cite{EREPR}.  It will be
fascinating to investigate this using the full string description of
this black hole.

Among the many questions worth exploring further, a
crucial one is to relate these horizon modes with the microscopic
degrees of freedom of a microcanonical description. If such a
correspondence can be constructed, perhaps then one can truly say that
we have a {\em complete} account of black hole entropy (not
unitarity). A discouraging observation in this context is that, at the
level of the gravity effective action, a Hawking-Page type transition
does not occur between the linear dilaton theory and the black
hole. This however, does not prevent us from exploring the issues
relating to the information paradox and black hole complementarity
using the black hole sigma model.


{\em Acknowledgements:}
I am grateful to Pravabati Chingangbam for discussions at an early
stage and for much subsequent motivation. I would like to acknowledge
HRI, Allahabad where some of this work was completed, and the
hospitality of various institutions where this work was presented at
various stages of development. CQUeST, Sogang University, Seoul, NTU
Taiwan, HIP Helsinki, IMSc Chennai, and IISc Bangalore.

\end{document}